# Excitonic Complexes and Emerging Interlayer Electron-Phonon Coupling in BN Encapsulated Monolayer Semiconductor Alloy: WS$_{0.6}$Se$_{1.4}$

Yuze Meng[1,2#], Tianmeng Wang[2#], Zhipeng Li[2,3], Ying Qin[4], Zhen Lian[2], Yanwen Chen[2], Michael C. Lucking[6], Kory Beach[6], Takashi Taniguchi[5], Kenji Watanabe[5], Sefaattin Tongay[4], Fengqi Song[1*], Humberto Terrones[6*], Su-Fei Shi[2,7*]

[1.] College of Physics, Nanjing University, Nanjing, 210093, P. R. China
[2.] Department of Chemical and Biological Engineering, Rensselaer Polytechnic Institute, Troy, NY 12180
[3.] School of Chemistry and Chemical Engineering, Shanghai Jiao Tong University, Shanghai, 200240, China
[4.] School for Engineering of Matter, Transport and Energy, Arizona State University, Tempe, Arizona 85287, USA
[5.] National Institute for Materials Science, 1-1 Namiki, Tsukuba 305-0044, Japan.
[6.] Department of Physics, Applied Physics, and Astronomy, Rensselaer Polytechnic Institute, Troy, NY 12180
[7.] Department of Electrical, Computer & Systems Engineering, Rensselaer Polytechnic Institute, Troy, NY 12180

[#] These authors contributed equally to this work
[*] Corresponding authors: shis2@rpi.edu, terroh@rpi.edu, songfengqi@nju.edu.cn

Abstract

Monolayer transition metal dichalcogenides (TMDs) possess superior optical properties, including the valley degree of freedom that can be accessed through the excitation light of certain helicity. While $WS_2$ and $WSe_2$ are known for their excellent valley polarization due to the strong spin-orbit coupling, the optical bandgap is limited by the ability to choose from only these two materials. This limitation can be overcome through the monolayer alloy semiconductor, $WS_{2x}Se_{2(1-x)}$, which promises an atomically thin semiconductor with tunable bandgap. In this work, we show that the high-quality BN encapsulated monolayer $WS_{0.6}Se_{1.4}$ inherits the superior optical properties of tungsten-based TMDs, including a trion splitting of ~ 6 meV and valley polarization as high as ~60%. In particular, we demonstrate for the first time the emerging and gate-tunable interlayer electron-phonon coupling in the BN/$WS_{0.6}Se_{1.4}$/BN van der Waals heterostructure, which renders the otherwise optically silent Raman modes visible. In addition, the emerging Raman signals can be drastically enhanced by the resonant coupling to the 2s state of the monolayer $WS_{0.6}Se_{1.4}$ A exciton. The BN/$WS_{2x}Se_{2(1-x)}$/BN van der Waals heterostructure with a tunable bandgap thus provides an exciting platform for exploring the valley degree of freedom and emerging excitonic physics in two-dimension.

Monolayer transitional metal dichalcogenides (TMDs), typically labelled as $MX_2$, where M is a transition metal atom (such as Mo or W), and X is a chalcogen atom (such as S, Se or Te), represents a new class of atomically thin semiconductor with superior optical properties such as the direct bandgap[1,2], large absorption, etc. TMDs also possess unique optical properties[3] which cannot be found in conventional semiconductors. Because of the reduced screening, the Coulomb interaction is enhanced in monolayer TMDs which gives rise to the strongly bound exciton and other higher order excitonic complexes such as trions[4-9]. The inversion symmetry breaking in TMDs also lifts the degeneracy at the corners of Brillouin zone, aka., K and K' valleys[10,11]. The valley degree of freedom[12] is similar to the spin and can be selectively excited through illumination with circularly polarized light[13]. The tungsten-based TMDs, such as $WS_2$ and $WSe_2$, retain high valley polarization due to the strong spin-orbit coupling[14-16], which induces a large splitting in the otherwise degenerate valence bands at K and K' points. However, the optical bandgap is limited by the ability to choose from only these two materials. This challenge can be overcome by the monolayer TMD alloy such as $Mo_xW_{1-x}Se_2$ or $WS_{2x}Se_{2(1-x)}$ which possess tunable optical bandgaps[17] depending on the composition ratio x. The capability of tuning the optical bandgap is critical for the valleytronics application of TMDs since the valley coherence time is longer as the excitation is close to the resonance of the optical bandgap. The tunable optical bandgap of the alloy TMDs, therefore, facilitate valleytronics applications when the excitation light wavelength is fixed. Despite the recent surge of interest on monolayer TMD alloys[17-31], the optical properties reported are limited by the quality of the sample and many questions remain open. Will the monolayer semiconductor alloy TMD still maintain the valley polarization? Will the unique trion physics and other optical properties of Tungsten based TMDs survive in the monolayer Tungsten based alloy TMD? Will the $BN/WS_{2x}Se_{2(1-x)}/BN$ van der Waals (vdW) heterostructure give rise to emerging interlayer electron-phonon coupling as in $BN/WSe_2/BN$ vdW heterostructure[32] and allow otherwise forbidden Raman modes to be visible? These are the questions that we will address in this work. In the present account, we will investigate the monolayer $WS_{0.6}Se_{1.4}$ as the model system, and we will examine if the monolayer $WS_{2x}Se_{2(1-x)}$ might inherit the strong spin-orbit coupling and thus the excellent valley polarization and the unique bandstructure which gives rise to dark exciton[33-38] in the monolayer $WS_2$ and $WSe_2$. We will address the abovementioned questions with a high-quality $BN/WS_{0.6}Se_{1.4}/BN$ heterostructure device.

The $BN/WS_{0.6}Se_{1.4}/BN$ heterostructure was constructed through a dry pickup method which ensures clean interface[9]. The carrier concentration can be controlled through an efficient top-gate, using the top BN layer as the gate dielectrics (Fig. 1a,b). Through low-temperature photoluminescence (PL) spectroscopy, we identify for the first time the clearly-resolved negative trion peaks in the monolayer alloy $WS_{0.6}Se_{1.4}$, including the intravalley and intervalley trions with a splitting of ~ 6 meV. This trion splitting is a result of the unique bandstructure of tungsten-based monolayer TMDs in which a spin-forbidden conduction band minimum (CBM) is the ground state, which has only been observed in high-quality monolayer $WS_2$ or $WSe_2$ samples. Employing PL spectroscopy with circularly polarized excitation, we also found that the valley polarization as high as ~ 60% in the $WS_{0.6}Se_{1.4}$ monolayer at 14 K, similar to what was reported in the high-quality monolayer $WSe_2$ and higher than that of the monolayer $WS_2$[7,16,39,40]. Even more strikingly, we found the unique interlayer electron-phonon coupling in the $BN/WS_{0.6}Se_{1.4}/BN$ vdW heterostructure, which is a sensitive function of the gate voltage. The emerging electron-phonon coupling reveals the otherwise optically forbidden Raman modes visible. And the emerging Raman signals can be enhanced by more than two orders of magnitude through the resonant coupling to a new electronic state in $WS_{0.6}Se_{1.4}$, which we reveal for the first time as the 2s state of the A exciton of monolayer alloy $WS_{0.6}Se_{1.4}$ by comparing with the mysterious X resonance in the previous work on $BN/WSe_2/BN$[32]. The vdW heterostructure of $BN/WS_{0.6}Se_{1.4}/BN$ thus provides an exciting platform for investigating emerging excitonic physics.

At room temperature, a strong PL centered at ~1.73 eV is observed from the monolayer $WS_{0.6}Se_{1.4}$, excited by the CW photoexcitation at 1.960 eV with the excitation power of 50 μW. The PL magnitude is similar to that of monolayer $WS_2$ or $WSe_2$ under the same excitation condition, confirming a direct bandgap nature. Compared with the room temperature PL peak at 2.01 eV for $WS_2$[6,7,42] and 1.65 eV for $WSe_2$[33], the PL of monolayer $WS_{0.6}Se_{1.4}$ is clearly tunable with the alloy composition. Previous work has shown that the PL spectrum peak position of the alloy can be tuned continuously from 1.65 eV (nearly pure $WSe_2$) to 1.97 eV (nearly pure $WS_2$) as the ratio of S varies from 0% to 100%[19,20,43,44]. Compared with the previous report of the PL peak as a function of the composition ratio, we determine that the ratio of S for our alloy $WS_{0.6}Se_{1.4}$ to be around 30%[19], consistent with our expectation (see Methods).

The Raman spectrum of the monolayer $WS_{0.6}Se_{1.4}$ is shown in Fig. 1c. In general, four main Raman modes can be observed in the range from 225 to 450 $cm^{-1}$, including $A_{1g(Se-W)}$ mode (259.7 $cm^{-1}$), $E_{2g(S-W)}$ mode (356.5 $cm^{-1}$), $A_{1g(S-W-Se)}$ mode (380.5 $cm^{-1}$) and $A_{1g(S-W)}$ mode (403.2 $cm^{-1}$), consistent with previous reports[19–22]. The co-existence of $A_{1g}$ mode for W-S and W-Se indicates that there is some random distribution of the S and Se[24], and the studied $WS_{0.6}Se_{1.4}$ is not a monolayer Janus alloy[26,45]. We also confirm this result through first-principles calculations with the $C_{3v}$ Janus structure, which shows four Raman active modes which we do not observe in our device (see SI).

Low-temperature PL at 14 K (Fig. 1d) demonstrates that, besides the exciton peak $X_0$, an additional sharp peak occurs which we attribute to the negative trion peak $X^-$. The trion binding energy is ~32 meV, similar to the reported values for other TMDs: 30 meV for $WSe_2$[14,46] and $MoSe_2$[5,8], and 34 meV for $WS_2$[6,7]. Despite the random distribution of the S and Se atoms[24], the in-plane inversion symmetry breaking remains for the monolayer $WS_{0.6}Se_{1.4}$, and the SHG measurement (Fig. 1e) clearly shows a six-fold symmetry, similar to that of monolayer TMDs[47–50]. Theoretical work on TMDs alloys has shown that they exhibit a strong SHG as the non-alloyed systems[51].

To explore the trion physics, we measured the PL spectra as a function of the top gate voltage. The gate voltage dependent PL spectra are plotted as a color plot in Fig. 2c, where the color represents the PL intensity. The line cuts, corresponding to PL spectra at specific gate voltages, are illustrated in Fig. 2a. It is evident from Fig. 2a that at the top gate voltage – 2 V, the exciton peak and the trion peak intensity are comparable, indicating that the $WS_{0.6}Se_{1.4}$ is lightly n-doped but is close to being intrinsic. As the top gate voltage increases, the electron concentration in the $WS_{0.6}Se_{1.4}$ increases. As a result, the PL intensity of trion increases while that of the exciton decreases. As the top gate voltage exceeds ~ 1 V, a new peak with more sensitive gate dependence occurs at 1.775 eV, and this peak redshifts to 1.757 eV at the top gate voltage of 1.75 V. Considering the thickness of the BN flake to be about 6.2 nm here, it would introduce electron density of about $3.4 \times 10^{12} cm^{-2}$ if we increase the gate voltage by 1 V from the charge neutral regime, based on the capacitance of 0.54 $\mu F/cm^2$ estimated from the geometry capacitance model. This sensitive change of PL at heavily n-doped regime has been previously reported in monolayer $WS_2$ and $WSe_2$, and has been recently assigned to the plasma coupled exciton[52–54].

Detailed examination of the trion peak shows that the trion peak is composed of two Lorentzian peaks. For example, at the gate voltage of 0.1 V (red curve in Fig. 2a), the trion PL can be fitted with two Lorentzian peaks centered at 1.775 eV and 1.782 eV (dashed lines in Fig. 2a). These two peaks originate from the exchange interaction induced energy splitting between the intravalley trion ($X_2^-$) and the intervalley trion ($X_1^-$). The detailed evolution of the two peaks as a function of the gate voltage can be found in the SI, and neither peak position is a sensitive function of the gate voltage. The separation between the two trions, obtained from the PL spectra, are ~ 6 meV. These two different trion configurations involve the exciton binding to one free electron in the spin-forbidden conduction band minimum (CBM), either in the same (intra) or the opposite (inter) valley. It is worth noting that the presence of the spin-forbidden conduction band at the K and K' valley is a characteristic of the band structure of the Tungsten based TMDs ($WS_2$ and $WSe_2$), but not the Molybdenum based TMDs. The fact we can observe the two split trion peaks suggests that monolayer alloy $WS_{0.6}Se_{1.4}$ inherits this characteristic from the Tungsten based TMDs. The splitting of the two trions peaks can also be resolved clearly in the reflectance measurement (Fig. 2b), and it is determined to be 6.3 meV in our monolayer $WS_{0.6}Se_{1.4}$ encapsulated with BN flakes. The splitting of the two trions has been reported to be 6-7 meV in $WSe_2$[39,55,56] and around 6 meV in $WS_2$[57] (also see SI). The similar splitting magnitude suggests similar exchange interaction in the monolayer alloy $WS_{2x}Se_{2(1-x)}$ to that in $WSe_2$ or $WS_2$.

We further investigate the valley polarization of the monolayer $WS_{0.6}Se_{1.4}$ by exciting the sample with the circularly polarized excitation and detecting the circularly polarized component of the emitted PL. Under right circularly polarized excitation ($\sigma^+$), the valley polarization, defined as the ratio of right circular and left circular components of the emitted PL, i.e., $P = \frac{I_{\sigma^+} - I_{\sigma^-}}{I_{\sigma^+} + I_{\sigma^-}}$, measures the capability of the TMD to maintain the valley information. As shown in Fig. 3a and 3b, it is evident that the $\sigma^+$ component of the PL intensity is stronger than that of the $\sigma^-$ component of the PL, for both the exciton and trion. This valley polarization can be clearly illustrated as the PL spectra at specific gate voltages, as shown in Fig.3d-f. The broad PL sideband centered at 1.70 eV, likely due to defects and phonon assisted PL, exhibits no noticeable difference in the $\sigma^+$ and the $\sigma^-$ component. However, the $\sigma^+$ component of the exciton or trion clearly exhibits selectivity. The valley polarization P can be quantitatively calculated using the integrated PL intensity, and its dependence on the gate voltage is shown in Fig. 3c. The valley polarization increases from ~ 30% at the top gate voltage of − 2 V to ~ 60% at the gate voltage 0.5 V, for both the exciton and trion. Compared with the valley polarization previously reported, 40% for $WS_2$[7,16] and 60% for $WSe_2$[14,40], the monolayer alloy $WS_{0.6}Se_{1.4}$ exhibits excellent valley polarization. The inheritance of the valley polarization in the monolayer alloy $WS_{0.6}Se_{1.4}$ is not surprising since the high valley polarization in $WS_2$ and $WSe_2$ originates from the large spin-orbit coupling in Tungsten which according to our DFT-PBE calculations is 0.429 eV for $WS_2$ and 0.466 eV for $WSe_2$ at the K point in the valence band maximum (VBM). Previous theoretical studies[58,59] have shown that the Bloch states of monolayer TMDs, symbolled as $MX_2$, is mostly dominated by the d orbital of the metal atom at the conduction band minimum (CBM) and valence band maximum (VBM), which locate at the corners of the Brillouin zone (K and K' points). The spin-orbit coupling leads to the large splitting of the VBM with the opposite spin configuration, which inhibits the intervalley scattering. As a result, Tungsten (W) based TMDs will better retain the valley polarization than the Molenbdeum (Mo) based TMDs because that the larger spin-orbit coupling in W will generate larger splitting in the VBM, causing the intervalley scattering energetically unfavorable for the same spin. We thus expect that the valley polarization in the monolayer alloy $WS_{0.6}Se_{1.4}$ is similar to that of $WSe_2$ or $WS_2$, despite the random distribution of the S and Se atoms. Our DFT-PBE computations carried out on $WS_{0.5}Se_{1.5}$ and $WS_1Se_1$ Janus confirms that, showing the spin-orbit coupling induced splittings of the VBM at K (K') points of 0.429 eV and 0.446 eV respectively (see SI).

Strikingly, vdW heterostructure based on TMDs exhibits emerging electron-phonon coupling[32,41], which we also found in the $BN/WS_{0.6}Se_{1.4}/BN$ heterostructure: otherwise optically silent Raman modes emerge through the resonant coupling to the electronic state of $WS_{0.6}Se_{1.4}$. The ZO phonon mode, which correspond to the out-of-plane vibration in BN, is originally forbidden but is brightened in the $BN/WS_{0.6}Se_{1.4}/BN$ heterostructure due to the different symmetry at the $BN/WS_{0.6}Se_{1.4}$ interface. The Raman signal is further drastically enhanced when the excitation is in resonance with the electronic state of $WS_{0.6}Se_{1.4}$. We reveal this by performing linearly polarized PL spectroscopy at 77 K. We use the linearly polarized laser for the excitation and measure the resulted PL signals either parallel or perpendicular (cross) to the polarization of the excitation. As shown in Fig. 4a, it is evident that additional peaks at 1.827 eV (I) and 1.859 eV (II) emerge for the parallel polarization configuration but not for the cross polarization. Furthermore, the new peaks' position changes as the photoexcitation energy changes. As shown in Fig. 4b, the peaks at 1.827 eV (I) and 1.859 eV (II) for the photoexcitation at 1.960 eV are shifted to 1.819 eV (I) and 1.851 eV (II) for the photoexcitation at 1.951 eV. However, the energy separation between emission photon energy and the excitation photon energy remains the same, 0.133 eV for peak II and 0.101 eV peak I, respectively. This is solid evidence that the two emerging peaks are Raman peaks at 1072 cm$^{-1}$ and 813 cm$^{-1}$. The assignment of the modes to the Raman peaks are also consistent with the sensitive polarization dependence as the Raman is a coherent process. Both Raman peaks are quenched by more two orders of magnitude and hardly visible (see SI) with the photoexcitation at 2.331 eV, suggesting that it is a resonant Raman process with one resonance close to 1.951-1.960 eV.

The Raman signal intensity closely correlates to the A exciton PL intensity of the monolayer $WS_{0.6}Se_{1.4}$. As shown in Fig. 4c, the Raman signals are the strongest for gate voltage between -2 to 0 V, in which A exciton PL is also the strongest, a signature for charge-neural $WS_{0.6}Se_{1.4}$. The two Raman peaks exhibit similar gate dependence, which is illustrated in Fig. 4d as the PL difference between the parallel and cross polarization detection configuration. Similar emerging resonant Raman signals have been reported previously in $BN/WSe_2/BN$ heterostructure[32,41], in which the forbidden hBN ZO mode (813 cm$^{-1}$) and hBN ZO mode + $WSe_2$ $A_{1g}$ (1072 cm$^{-1}$) modes couples resonantly to the A exciton of $WSe_2$ and a mysterious electronic state X. The existence of the state X can be observed from the differential reflectance ($\Delta R/R$) spectra as a function of the gate voltage. We thus perform the differential reflectance measurements on the $BN/WS_{0.6}Se_{1.4}/BN$ heterostructure, as shown in Fig. 5a. It is clear that besides the obvious A exciton ($X_0$, ~1.80 eV), there is a subtle feature around 1.945 eV which disappears as the gate voltage increased to 0.75 V. This X* is not due to the B exciton resonance, since the B exciton resonance is ~400 meV below the A exciton resonance due to the large spin-orbit coupling (see SI for the reflectance spectra showing the B exciton resonance). To better illustrate the change of the differential reflectance spectra as the function of the gate voltage, we take the derivative of the $\Delta R/R$ spectra which we show in Fig. 5b and 5c. It is evident that we observe an emerging feature at 1.945 eV in the gate voltage range between -2 to 0 V. This new absorption resonance (X*) is close to the resonant photoexcitation of the emerging Raman peaks. It is worth noting that although the emerging absorption peak (X*) position (1.945 eV) is different from the X peak (1.85 eV) in the BN/WSe$_2$/BN structurec[32], the energy difference between the X* and the A exciton (X$_0$) in WS$_{0.6}$Se$_{1.4}$ is ~140 meV, similar to the energy difference between the X peak and A exciton energy of WSe$_2$ (~ 0.14 eV)[32]. Considering the similar Coulomb interaction in BN encapsulated monolayer WS$_{0.6}$Se$_{1.4}$ and BN encapsulated monolayer WSe$_2$, we conclude that the newly observed absorption peak is the 2s states of A exciton in WS$_{0.6}$Se$_{1.4}$ and the mysterious X state observed previously is the 2s state of the A exciton of WSe$_2$[60,61]. This assignment is consistent with the gate dependence of the absorption spectra (Fig. 5d), since the 2s state only exists in a charge-neutral region. This assignment of X is also consistent with the recent B field dependence study of higher order excited states in WSe$_2$, which reported 2s state at the similar energy as the X state[62,63]. We thus report for the first time the observation of the 2s state of the monolayer WS$_{0.6}$Se$_{1.4}$.

In summary, we have encapsulated a monolayer alloy WS$_{0.6}$Se$_{1.4}$ with BN flakes and fabricated high-quality devices with an efficient top gate. As expected from the composition ratio, the monolayer WS$_{0.6}$Se$_{1.4}$ exhibits distinct optical bandgap compared with that of the monolayer WS$_2$ or WSe$_2$. However, all the superior optical properties such as the valley selectivity and unique trions physics are retained in the monolayer alloy WS$_{0.6}$Se$_{1.4}$. Negative trion peaks splitting of ~ 6 meV is also observed. In particular, we found the emerging interlayer electron-phonon when the forbidden phonon modes in BN resonantly couple to the 2s state of the A exciton of the monolayer WS$_{0.6}$Se$_{1.4}$. With the tunability of the composition ratio, BN encapsulated monolayer TMD alloy with controllable optical bandgap provides an exciting playground for tunable excitonic optics and emerging electron-phonon interactions.

**Material synthesis:** WS$_{2x}$Se$_{2(1-x)}$ crystals were synthesized at x=0.3 (i.e., 30% S and 70% Se) composition using chemical vapor transport technique using elemental tungsten (Sigma Aldrich, 99.9998%), sulfur (Sigma Aldrich, 99.9999%), and selenium (Alfa Aesar, 99.9999%) as precursors. These precursors were mixed and loaded into pre-cleaned and heated quartz ampoules with the x values kept at 0.3. The ampule was then evacuated (to less than $10^{-5}$ Torr), sealed, and placed in a two-zone furnace. The hot zone of the furnace was kept at 1050 ± 1 °C with a temperature difference of 3.0 ± 0.1 °C/cm over a distance of 21 cm. After 28 days, the ampule was naturally cooled to room temperature, and crystals were found at the interior wall of the ampoule. Samples were characterized through SEM, EDS, Raman, and XRD. We note that even substantial changes in the precursor x composition still resulted in crystals with mostly x=0.3 values which suggests that in the ternary phase diagram there is a strong energetic minimum to preferentially stabilize this particular composition.

**Device preparation and characterization:** The BN encapsulated monolayer $WS_{0.6}Se_{1.4}$ was constructed through the dry pickup process[64,65] (see SI for details), and the van der Waals (vdW) heterostructure was constructed by using the BN stamp to pick up each constituent layer. The final vdW structure was deposited on the silicon chip with 300 nm thermal oxide, as shown in the schematic in Fig. 1b. During the process, the monolayer $WS_{0.6}Se_{1.4}$ was never exposed to any polymer, which is critical to ensure the quality of the PL spectra with narrow peaks, and we exploit the as-prepared monolayer $WS_{0.6}Se_{1.4}$ to probe the intrinsic property of the monolayer TMD alloy. One piece of few-layer graphene, labeled as graphite, was used as the contact electrode and the second piece of few-layer graphene was used as the transparent top gate electrode, enabling both optical access and efficient tunability of the carrier concentration in the monolayer $WS_{0.6}Se_{1.4}$. The optical spectroscopy measurements were performed in a home-built confocal microscope setup, with the focus spot size of ~ 2 μm. Continuous wave (CW) laser centered at 633 nm or 532 nm was used as the excitation source for the PL and Raman measurements. A supercontinuum white laser (Fianium) was used as the light source for the reflectance measurement. The second harmonic generation (SHG) was performed with a femtosecond pulsed laser (pulse width ~ 120 fs) as the excitation source, generated from a Ti: sapphire oscillator (Coherent) with the repetition rate of 80 MHz. The low-temperature measurement was performed with a closed-cycle optical cryostat which has the base temperature of ~ 14 K.

## Supporting information

Details about sample preparation, optical characterization and data analysis can be found in the supporting information (SI).

## Conflicts of interest

There are no conflicts of interest to declare.

## Acknowledgments

The device fabrication was supported by Micro and Nanofabrication Clean Room (MNCR) at Rensselaer Polytechnic Institute (RPI). S.-F. Shi acknowledges support from the AFOSR through Grant FA9550-18-1-0312. HT, MCL and KB acknowledge support of: the NSF grant EFRI-1433311, the supercomputer time was provided by the Center for Computational Innovations (CCI) at RPI and the Extreme Science and Engineering Discovery Environment (XSEDE, project TG-DMR17008), which is supported by National Science Foundation grant number ACI-1053575. Y. M. acknowledges the support from China Scholarship Council. L. Zhen acknowledges support by the NY State Empire State Development's Division of Science, Technology and Innovation (NYSTAR) through Focus Center-NY–RPI Contract C150117.

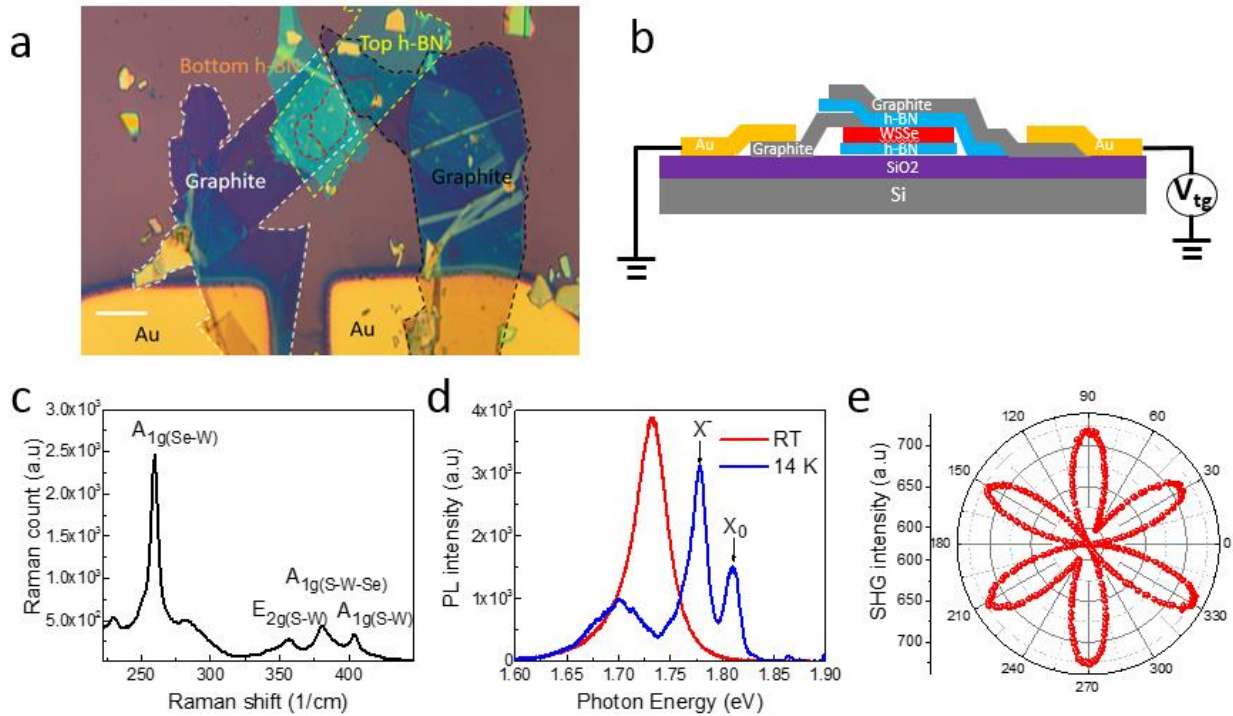

**Figure 1. Optical spectra of the monolayer WS$_{0.6}$Se$_{1.4}$.** (a) Optical microscope image of the BN encapsulated WS$_{0.6}$Se$_{1.4}$ monolayer with one few-layer graphene (graphite) acting as the contact electrode and one few-layer graphene working as the top gate electrode. The monolayer WS$_{0.6}$Se$_{1.4}$ is outlined with the red dashed line. Scale bar: 20 μm. (b) A schematic side view of the BN encapsulated WS$_{0.6}$Se$_{1.4}$ monolayer device. (c) Raman spectrum of the WSSe monolayer at room temperature, with a CW laser centered at 532 nm (2.331 eV) being the excitation source. (d) PL spectra of the WS$_{0.6}$Se$_{1.4}$ monolayer at room temperature (red) and 14 K (blue), with a CW laser centered at 633 nm (1.959 eV). (e) Second harmonic generation (SHG) signal as a function of the polarization angle between the optical field and the monolayer WS$_{0.6}$Se$_{1.4}$ obtained at room temperature, with a pulsed laser (width ~ 120 fs) centered at 800 nm (1.55 eV) being the excitation source.

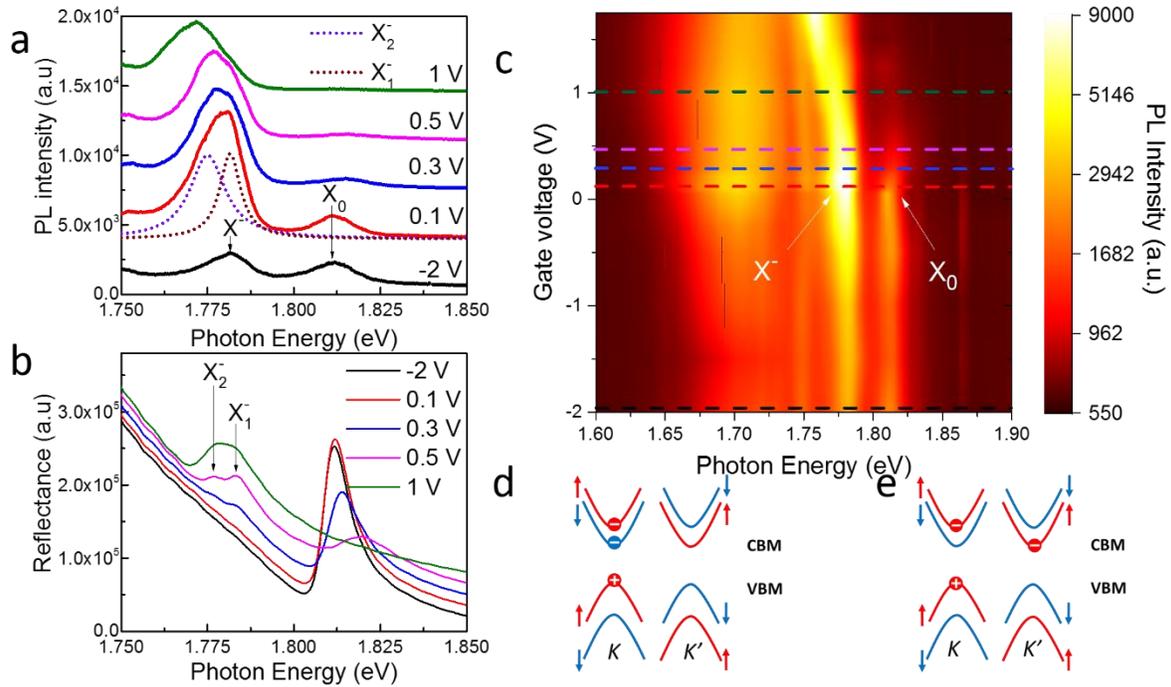

**Figure 2. Gate-voltage dependence of the PL spectra of the monolayer $WS_{0.6}Se_{1.4}$.** (a) The PL spectra of $WS_{0.6}Se_{1.4}$ monolayer at different gate voltages at 14 K. $X_0$: exciton. $X^-$: trion. The trion peak can be decomposed of two trion peaks (dashed lines). (b) The reflectance spectra of $WS_{0.6}Se_{1.4}$ monolayer at different gate voltages at 14 K. $X_1^-$: inter-valley trion. $X_2^-$: intra-valley trion. (c) Color plot of the $WS_{0.6}Se_{1.4}$ monolayer PL spectra as a function of the gate voltage. The color represents the PL intensity. The dashed lines correspond to the spectra at specific gate voltage as shown in (a). (d) and (e) are schematic of intravalley and intervalley trions, respectively.

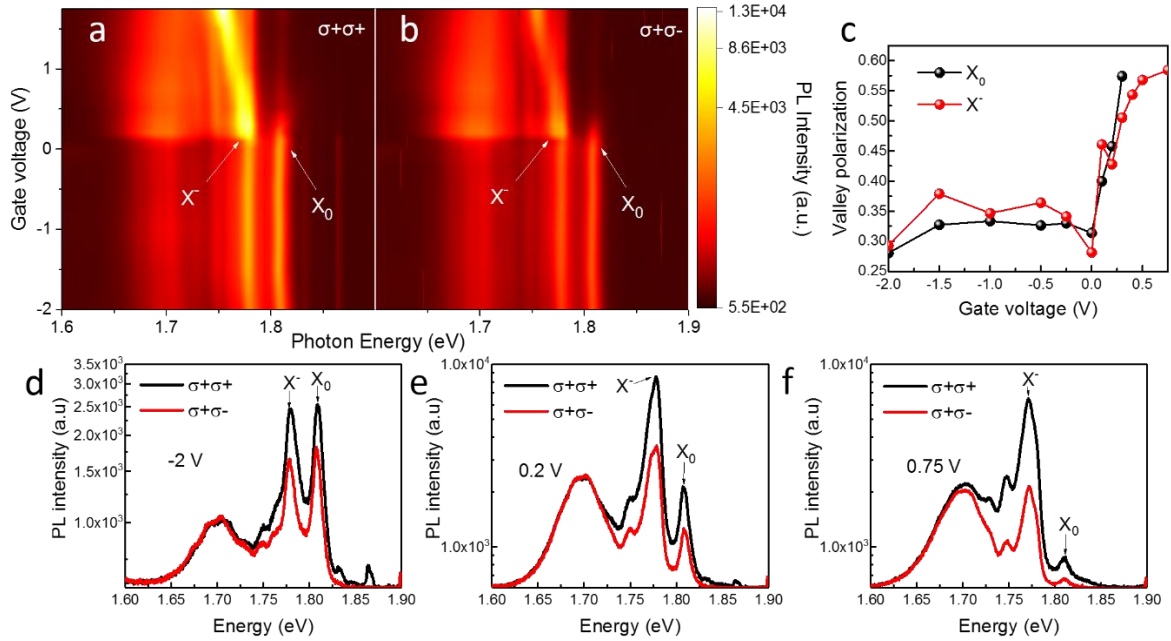

**Figure 3. Valley polarized PL spectra of monolayer WS$_{0.6}$Se$_{1.4}$.** (a) and (b) are the color plot of the PL spectra as a function of the gate voltage for the $\sigma^+\sigma^+$ (a) and $\sigma^+\sigma^-$ (b) configuration, correponding to $\sigma^+$ excitation and $\sigma^+$ (a) or $\sigma^-$ (b) detection. The color represents the PL intensity, with the same color scale bar. (c) The calculated valley polarization P, defined as $\frac{I_{\sigma^+}-I_{\sigma^-}}{I_{\sigma^+}+I_{\sigma^-}}$, is plotted as a function of the gate voltage for both the exciton and trion. (d)-(f) PL spectra with circular excitation $\sigma^+$ are detected with same ($\sigma^+$) and opposite ($\sigma^-$) helicity for gate voltage of (d) -2 V, (e) 0.2 V and (f) 1.75 V at 14 K.

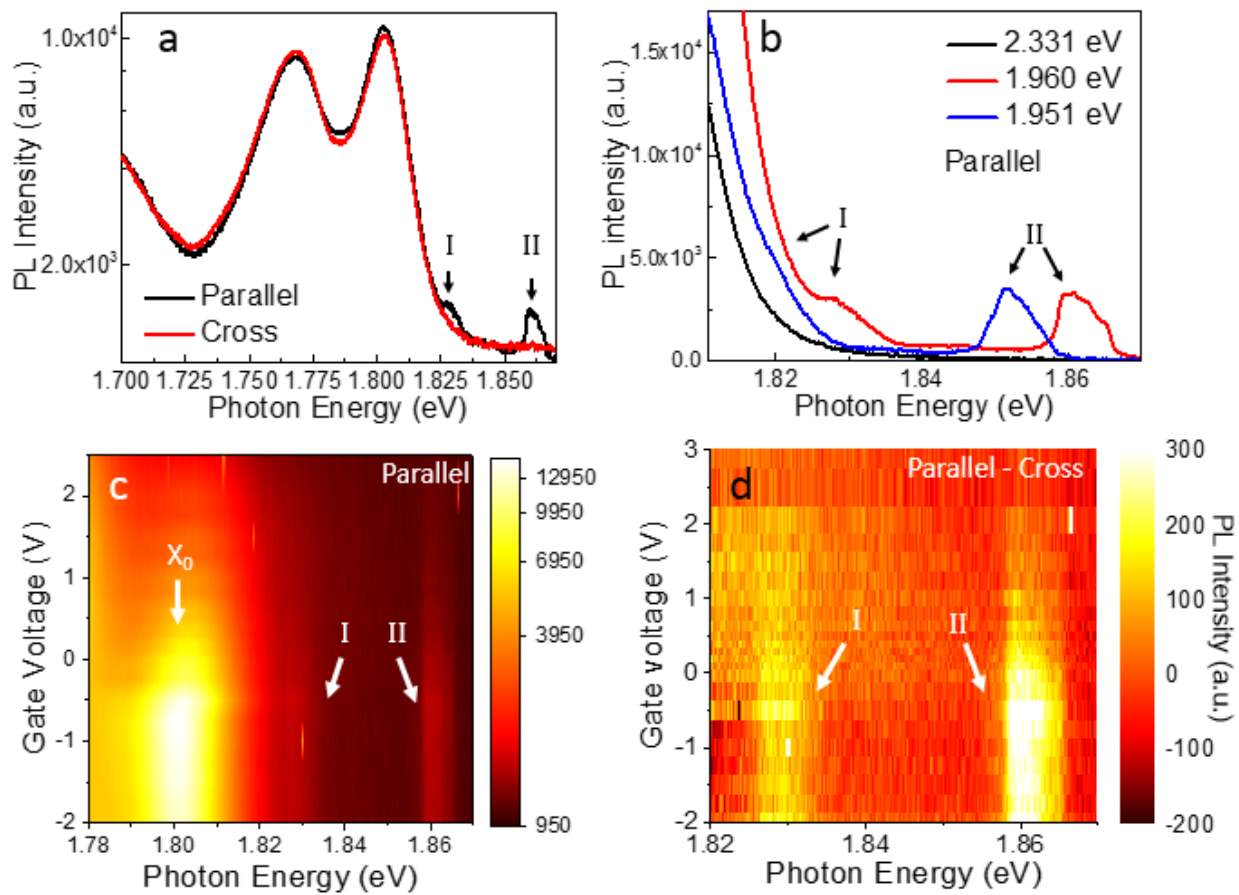

**Figure 4. Emerging interlayer electron-phonon coupling in the BN/ WS$_{0.6}$Se$_{1.4}$/BN vdW heterostructure.** (a) PL spectra of the monolayer WS$_{0.6}$Se$_{1.4}$ with linearly polarized photoexcitation of CW laser centered at 1.960 eV, detected by a parallel and perpendicular (cross) polarizer at 77 K. (b) PL spectra of the monolayer WS$_{0.6}$Se$_{1.4}$ with CW excitation at different photon energy: 2.331eV (black), 1.959 eV (red) and 1.953eV (blue) at 77 K. All spectra are taken with the parallel configuration. (c) PL spectra of the monolayer WS$_{0.6}$Se$_{1.4}$ in the parallel configuration as a function of the gate voltage at 77 K. The color represents the PL intensity. (d) PL spectra difference between the parallel and cross configuration as a function of the gate voltage at 77 K. The color represents the PL intensity difference of the parallel and cross configuration. Spectra in (c) and (d) are taken with the CW laser excitation with photon energy centered at 1.960 eV.

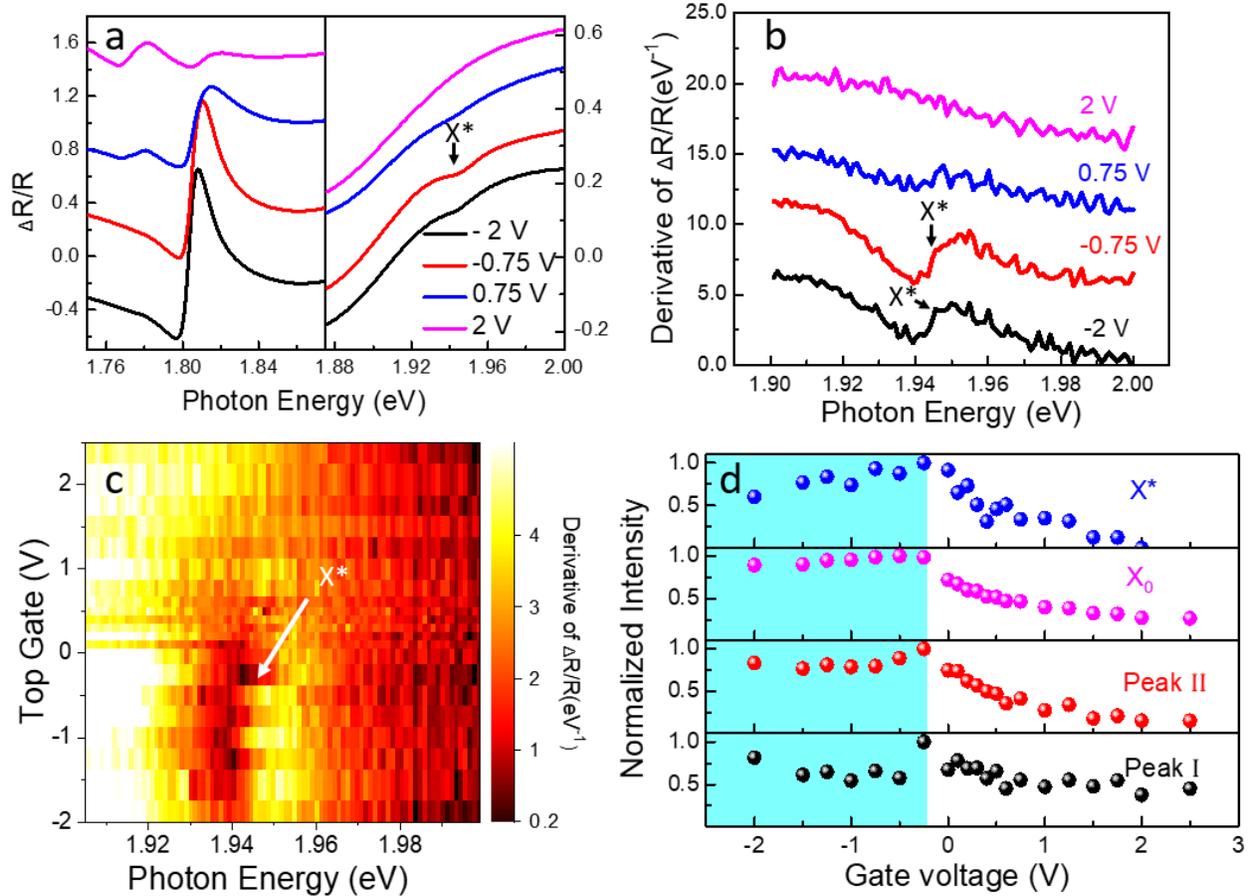

**Figure 5. Gate-voltage dependent differential reflectance spectra** (a) Differential reflectance spectra the BN encapsulated monolayer WS$_{0.6}$Se$_{1.4}$ for different gate voltages. (b) Derivative of the differential reflectance (DR) spectra of the monolayer WS$_{0.6}$Se$_{1.4}$ around 1.94 eV as a function of the gate voltage at 14 K. The color represents the derivative of the $\Delta R/R$ signal. (c) The derivative of the corresponding differential reflectance spectra shown in (a). X* indicates where the new feature occurs. (d) Normalized PL intensity of the peaks centered at 1.83 eV (black dots), 1.86 eV (red dots), A exciton (magenta dots) and normalized derivative of $\Delta R/R$ intensity of the new resonance feature in (b) near 1.94 eV (blue dots) as a function of the gate voltage. The light blue color marks the region where the monolayer WS$_{0.6}$Se$_{1.4}$ is nearly intrinsic.

TOC:

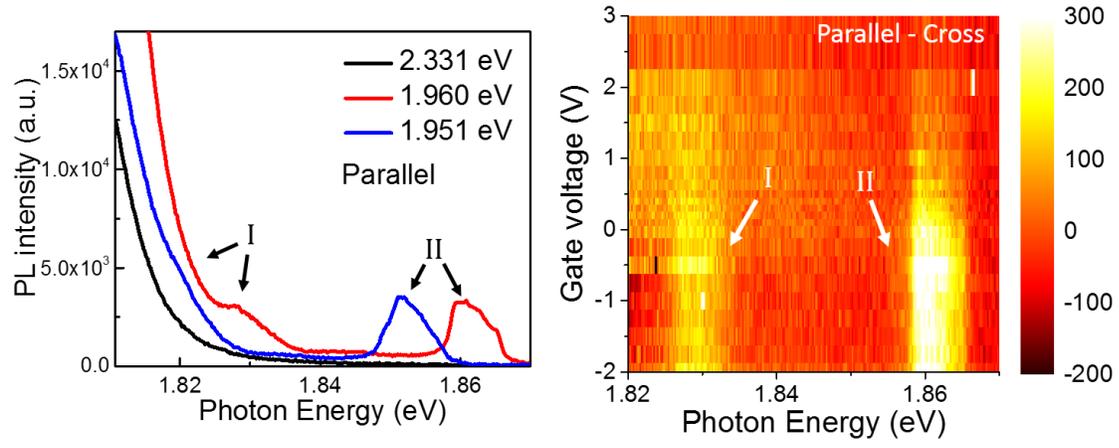